\documentclass[12pt]{article}
\usepackage[dvips]{graphicx}
\usepackage{pdproc}
\newcommand{\beq}{\begin{eqnarray}}
\newcommand{\eeq}{\end{eqnarray}}

  \makeatletter 
  \def\@cite#1{[#1]} 
  \makeatother    
\begin{document}

\renewcommand{\thefootnote}{\alph{footnote}}

\title{
 New Conformal Field Theories with Anomalous Dimensions
}

\author{ ETSUKO ITOU}

\address{ Department of Physics, Graduate School of Science, Osaka University,\\Toyonaka, Osaka 560-0043, Japan
\\ {\rm E-mail: itou@het.phys.sci.osaka-u.ac.jp}}

\abstract{
We find a class of fixed point theory for 2- and 3-dimensional non-linear sigma models using Wilsonian renormalization group (WRG) approach.
In $2$-dimensional case, the fixed point theory is equivalent to the Witten's semi-infinite cigar model.
In $3$-dimensional case, the theory has one parameter which describes a marginal deformation from the infrared to ultraviolet fixed points of the $CP^N$ model in the theory spaces.
}

\normalsize\baselineskip=15pt

\section{Introduction}

The Wilsonian renormalization group equation describes the variation of effective action $S[\Omega; t]$ when the cutoff energy scale $\Lambda$ is changed to $\Lambda (\delta t)=\Lambda e^{-\delta t}$ in $D$ dimensional field theory \cite{Wilson Kogut}\cite{Wegner and Houghton}\cite{Morris}:
\beq  
\frac{d}{dt}S[\Omega; t]&=&\frac{1}{2\delta t} \int_{p'} tr \ln \left(\frac{\delta^2 S}{\delta \Omega^i \delta \Omega^j}\right)
-\frac{1}{2 \delta t}\int_{p'} \int_{q'} \frac{\delta S}{\delta \Omega^i (p')} \left(\frac{\delta^2 S}{\delta \Omega^i (p')\delta \Omega^j (q')} \right)^{-1} \frac{\delta S}{\delta \Omega^j (q')} \nonumber\\
&&+ \left[D-\sum_{\Omega_i} \int_p \hat{\Omega}_i (p) \left(d_{\Omega_i}+\gamma_{\Omega_i}+\hat{p}^{\mu} \frac{\partial}{\partial \hat{p}^{\mu}} \right) \frac{\delta}{\delta \hat{\Omega}_i (p)} \right] \hat{S},\label{WRG-1} 
\eeq
where $d_{\Omega}$ and $\gamma_{\Omega}$ denote the canonical and anomalous dimensions of the field $\Omega$.
The caret indicates dimensionless quantities.
The first and second terms in eq.(\ref{WRG-1}) correspond to the one-loop and tree diagrams contributions, when internal of fields with high momentum $\Lambda (\delta t)<p<\Lambda$ lines are eliminated. The remaining terms come from the rescaling of fields to normalize the coefficient of the kinetic term to unity.
We impose ${\cal N}= 2$ supersymmetry on the action and consider only K\"{a}hler potential term to define the ${\cal N}=2$ supersymmetric nonlinear sigma model in two- and three- dimensions
\beq
S&=&\int d^2 \theta d^2 \bar{\theta} d^D x K[\Phi, \Phi^\dag]\nonumber\\
&=&\int d^D x \Bigg[g_{n \bar{m}}\left(\partial^{\mu} \varphi^n \partial_{\mu} \varphi^{* \bar{m}} +\frac{i}{2} \bar{\psi}^{\bar{m}} \sigma^{\mu}(D_{\mu} \psi)^n +\frac{i}{2} \psi^{n} \bar{\sigma}^{\mu}(D_{\mu} \bar{\psi})^{\bar{m}} +\bar{F}^{\bar{m}} F^{n}\right) \nonumber\\
&&-\frac{1}{2} K_{,nm \bar{l}} \bar{F}^{\bar{l}} \psi^n \psi^m -\frac{1}{2} K_{,n \bar{m} \bar{l}} F^{n} \bar{\psi}^{\bar{m}} \bar{\psi}^{\bar{l}}+\frac{1}{4} K_{,nm \bar{k} \bar{l}} (\bar{\psi}^{\bar{k}} \bar{\psi}^{\bar{l}})(\psi^n \psi^m)\Bigg],\label{action}
\eeq
where $\Phi^n$ denote chiral superfields, whose components fields are complex scalars $\varphi^n(x)$, Dirac fermions $\psi^n(x)$ and complex auxiliary fields $F^n(x)$. The K\"{a}hler metric of the target space $g_{i \bar{j}}$ is given by the K\"{a}hler potential
$g_{i \bar{j}}(\varphi, \varphi^{*})
=\frac{\partial^2 K(\varphi, \varphi^{*})}{\partial \varphi^i \partial \varphi^{* \bar{j}}}.$

From the WRG equation, the $\beta$ function for the K\"{a}hler metric is given by \cite{SU2dim}
\beq
\beta(g_{i \bar{j}})&=&\frac{1}{2 \pi}R_{i \bar{j}} +\gamma \Big[\varphi^k g_{i \bar{j},k} +\varphi^{* \bar{k}}g_{i \bar{j},\bar{k}}+2g_{i \bar{j}} \Big]+d_{\varphi}\Big[\varphi^k g_{i \bar{j},k} +\varphi^{* \bar{k}}g_{i \bar{j},\bar{k}} \Big].\label{beta}
\eeq
Note that our $\beta$ function reduces to the Ricci tensor when the anomalous dimension of the fields vanishes. The second term, proportional to the anomalous dimension $\gamma$, which is not reparametrization invariant because of the renormalization condition of the fields breaks reparametrization invariance.
Since the K\"{a}hler metric contain the infinite number of coupling constants, the above WRG equation is an infinite set of differential equations for these coupling constants.

\section{Fixed points with $U(N)$ symmetry in two-dimensions}
Let us derive the action of the conformal field theory 
corresponding to the fixed-point of the $\beta$ function (\ref{beta}) for $d_{\varphi}=0$.
Since Ricci curvature $R_{i\bar{j}}$ is a second derivative of the metric $g_{i \bar{j}}$, the equation is a set of coupled partial differential equations, and is very difficult to solve in general.
So we simplify the problem by assuming symmetry ${\bf U}(N)$ for the K\"{a}hler potential.
\beq
K[\varphi,\varphi^\dag]&=&\sum^{\infty}_{n=1} g_n x^n \equiv f(x)\label{SUN}
\eeq
where $x$ is the ${\bf U}(N)$ invariant combination $x \equiv \vec{\varphi} \cdot \vec{\varphi}^\dag$
of the $N$ components scalar fields $\vec{\varphi}=(\varphi^1,\varphi^2,\cdots,\varphi^N)$. The coefficients $g_n$ play the role of an infinite number of coupling constants which depend on the cutoff scale $t$.

We substitute the metric and Ricci tensor, obtained by the above K\"ahler potential, for the $\beta$ function (\ref{beta-2dim}), and find that the fixed point theory satisfies following differential equation
\beq
\frac{\partial}{\partial t}f'&=&\frac{1}{2\pi}\Big[(N-1)\frac{f''}{f'} +\frac{2f''+f''' x}{f'+f'' x}  \Big]- 2\gamma(f'+f''x)\label{f'}=0,\label{beta-f}
\eeq
where $f'=\frac{df}{dx}$.

We found that the solution of this equation has a free parameter corresponding to the anomalous dimension of the field. 
In fact, the solution is very simple when the target manifold is of complex one-dimension $N=1$:
\beq
f'=\frac{1}{ax} \ln (1+ax).
\eeq
Here, $a \equiv -4 \pi \gamma$.
This $f'$ gives the metric of the target space
\beq
g_{i \bar{j}}=f'+f''x=\frac{1}{1+ax}.\label{CFT-metric}
\eeq
\begin{figure}[htb]
\begin{center}
\includegraphics*[width=5cm]{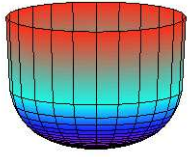}
\caption{%
The target manifold embedded in $3$-dimensional flat Euclidean spaces.
}
\label{fig1}
\end{center}
\end{figure}
Note that the metric has only one component, and the indices $i$ and $\bar{j}$ is $1$.
The volume and the distance from the origin ($|x|=0$) to infinity ($|x|=\infty$) of target spaces are divergent, while the length of the circumference at the infinity is finite. Therefore, the shape of the target manifold is a semi-infinite cigar with radius $\sqrt{\frac{1}{a}}$.

This solution has been discussed in other context \cite{cigar}.
They consider the non-linear sigma model coupled with dilaton on curved space-time.
The perturbative $1$-loop $\beta$ function for K\"ahler metric has the contribution of the dilaton field,
\beq
\beta (g_{i \bar{j}})=R_{i \bar{j}}+ 2 \nabla_i \nabla_{\bar{j}} \Phi, \nonumber
\eeq
where $\Phi$ denotes the dilaton field.
That last term plays the same role as the anomalous dimension term in (\ref{beta}).
In fact, the above non-trivial dilaton gradient in space-time is equivalent to assigning a non-trivial Weyl transformation law to target space coordinates. 
Then, we can obtain the same fixed point theory using WRG equation.

\section{Three-dimensional case}

Next, we consider $3$-dimensional non-linear sigma model.
It is nonrenormalizable within the perturbative method, so we need non-perturbative analysis \cite{SU3dim}.

For example, we consider $CP^N$ model which is defined by following K\"ahler potential:
\beq
K[\Phi,\Phi^\dag]=\frac{1}{\lambda^2} \ln (1+\Phi^i \Phi^{\dag \bar{j}}).
\eeq
where $i=1,\cdots,N$.
Using the $\beta$ function (\ref{beta}) for $d_{\varphi}=\frac{1}{2}$, We obtain the anomalous dimension of scalar fields (or chiral superfields) and $\beta$ function of the coupling constant $\lambda$ as
\beq
\gamma&=&- \frac{(N+1) \lambda^2}{4\pi^2},  \hspace{1cm}
\beta(\lambda)=-\frac{N+1}{4\pi^2}\lambda^3+\frac{1}{2} \lambda .
\eeq
We find there are an IR fixed point at $\lambda=0$ and an UV fixed point at $\lambda^2=\frac{2 \pi^2}{N+1}\equiv \lambda_c^2$.

Now, we investigate the conformal field theories defined as the fixed point of the $\beta$ function (\ref{beta}) for $d_{\varphi}=\frac{1}{2}$.
To simplify, we assume ${\bf S}{\bf U}(N)$ symmetric K\"{a}hler potential (\ref{SUN}) as before.
We substitute the metric and Ricci tensor for the $\beta$ function (\ref{beta}) and obtain the following differential equation.
\beq
\frac{\partial}{\partial t}f'=\frac{1}{2\pi^2}\Big[(N-1)\frac{f''}{f'} +\frac{2f''+f''' x}{f'+f'' x}  \Big]- 2\gamma(f'+f''x)\label{f'}-f''x =0\label{beta=0-3dim}.
\eeq

To obtain a conformal field theory, we must solve the differential equation
The function $f(x)$ is a polynomial of infinite degree, and it is hard to solve it analytically. So we expand the function $f(x)$ and the equation (\ref{beta=0-3dim}) around $x \approx 0$. 
Then the following function satisfies $\beta=0$ for any values of parameter $g_2$:
\beq
f'&=&1+2g_2 x + [\frac{2(3N+5)}{N+2}g_2^2 +\frac{2 \pi^2 }{N+2}g_2]x^2 \nonumber\\
&&\hspace{-0.9cm}+\frac{4}{3(N+2)(N+3)}[(16N^2 +66N+62)g_2^3 +2\pi^2 (6N+14)g_2^2 +2\pi^4 g_2  ]x^3+\cdots.\label{3-dim-sol}
\eeq
The parammeter $g_2$ corresponds to the anomalous dimension of scalar field as in the $2$-dimensional case.
The function $f(x)$ describe the conformal field theory and has one free parameter $g_2$.
Thus, if we fix the value of $g_2$, we obtain a conformal field theory.

For example, we take $
g_2=-\frac{1}{2} \cdot \frac{2\pi^2}{N+1} \equiv -\frac{1}{2}a.
$
Then, the function $f(x)$ becomes 
\beq
f(x) =\frac{1}{a } \ln (1+ a x ),\label{CP-UV}
\eeq
and this is the K\"{a}hler potential of C$P^N$ model.
Thus one of the novel ${\bf SU}(N)$ symmetric conformal field theory is equal to the UV fixed point theory of C$P^N$ model for the specific value of $g_2$.
In this case, the symmetry of this theory enhances to ${\bf SU}(N+1)$ because the C$P^N$ model has the isometry ${\bf SU}(N+1)$.

\section{Summary}

 We constructed a class of the ${\bf SU}(N)$ symmetric conformal field theory by using the WRG equation. This has one free parameter corresponding to the anomalous dimension of the scalar fields. If we choose a specific value of the parameter, we recover the conformal field theory defined at the UV fixed point of C$P^N$ model and the symmetry is enhanced to ${\bf SU}(N+1)$.

\section{Acknowledgements}
\vspace{-0.3cm}
We would like to thank K. Higashijima, K. Hori, Y. Kitazawa and J. Nishiumura for useful discussions.
This work was supported by Japan Society for Promotion of Science.
\vspace{-0.3cm}

\bibliographystyle{plain}

\end{document}